\documentclass[nofootinbib]{revtex4}
\usepackage[english]{babel}
\usepackage{array,booktabs}
\usepackage{array} 
\usepackage{lipsum}   
\usepackage{calc}
\usepackage{pdflscape}
\usepackage{color}
\usepackage{float}
\usepackage[font=small,labelfont=bf]{caption}
\usepackage{graphicx}
\usepackage{amsmath}
\usepackage[nodisplayskipstretch]{setspace}

\usepackage{setspace}
\usepackage{tabularx}

\usepackage{float}
\usepackage{color}
\usepackage{amsmath}
\usepackage{float}
\usepackage{calc}
\usepackage{pdflscape}
\usepackage{color}
\usepackage{float}
\usepackage{subfigure}
\usepackage[font=small,labelfont=bf]{caption}
\usepackage{graphicx}
\begin{document}{\setlength\abovedisplayskip{4pt}}

\title{Pomeron inspired Neutrino-nucleon Cross sections at Ultra High Energy}
\author{Kalpana Bora\footnote{E-mail: kalpana.bora@gmail.com}, Neelakshi Sarma\footnote{E-mail: nsarma25@gmail.com}, Jaydip Singh\footnote{E-mail: jaydip.singh@gmail.com}} 
\affiliation{Department Of Physics, Gauhati University, $Guwahati-781014^{\ast,\dagger}$ \\ Department Of Physics, Lucknow University, $Lucknow^{\ddagger}$}

\begin{abstract}
Studies on neutrino-nucleon ($\nu N$) cross sections at different energy scales have regained interest due to increasing importance of precision measurements, as they are needed as an ingredient in all neutrino experiments. In this paper we have calculated both charged current (CC) and neutral current (NC) $\nu$N scattering cross sections at Ultra High Energy (UHE) regime in the neutrino energy ($E_{\nu}$) region i.e. $10^{9} GeV \le E_{\nu} \le 10^{12}$ GeV using QCD inspired double asymptotic limit fit of electron-proton structure function $F_{2}^{ep}$ to low $\mathit{x}$ HERA data. The form $F_{2}^{ep} \sim x^{-\lambda(Q^{2})}$ used in our analysis, can be conjectured like a  dynamic pomeron (DP)-type behaviour.  We also find an analytic form of the total cross sections, $\sigma_{CC}^{\nu N}$ and $\sigma_{NC}^{\nu N}$ which appear to be of hard-pomeron exchange types. A comparative analysis  of our results with those available in literature is also done. An improved understanding of $\nu N$ interactions at UHE are essentially important for future oscillation experiments. Future measurements will support/confront our predictions.   
\end{abstract}
\maketitle

\section{Introduction}	
\setlength{\baselineskip}{13pt}
 Neutrino nucleon scattering cross sections play a pivotal role in all neutrino oscillation experiments. Such experiments make use of neutrinos coming from natural resources as well as from artificial (man-made) resources \cite {1}. In any neutrino experiments, neutrinos are scattered off a nucleon/nucleus of the detector. Number of events (of signal process) are observed experimentally, which is proportional to the flux of the incoming neutrinos, cross section and probability of the signal process. Neutrinos coming to the earth from natural sources have their origin in the sun, active galactic nuclei (AGN) and core of supernovae-they are believed to play crucial role in various astrophysical phenomenon. The information obtained from astrophysical objects and mechanisms is complimentary to that available from electromagnetic or hadronic interactions. Neutrino interactions across various energy scales can be categorised as  \cite {1}:\\

(i) Thresholdless process : Here the neutrino interaction energy is in the range 0 - 1 MeV. Both coherent scattering and neutrino capture on radioactive nuclei (enhanced or stimulated beta decay emission) fall under thresholdless process.\\

(ii) Low energy nuclear process : Here the neutrino energy scale is from 0 - 100 MeV. At such process, it is possible to probe the target nucleus at smallest length scales.\\

(iii) Intermediate energy process : The energy scale for intermediate energy process is from 0.1 - 20 GeV, where the neutrino scattering becomes more diversed and complicated. At these energies neutrino elastically scatter off an entire nucleon from target nucleus. Both charged current (also called quasi-elastic, QE) and neutral current (elastic) neutrino scattering processes fall under this process. \\
 
(iv) High energy process : The energy range of high energy process is from 20 - 500 GeV. This process includes Deep inelastic scattering (DIS) where neutrino can resolve individual quark constituents of the nucleon.\\

(v) Ultra high energy (UHE) process : The energy range of UHE process is from 0.5 TeV - 1 EeV. In recent times, highest energy neutrino recorded is around $\sim$ PeV, \cite {2} which indeed provides many opportunities for researchers to work with UHE neutrinos coming from astrophysical resources \cite {3}. \\

   Neutrino DIS processes have been used to validate the standard model (SM) and also probe nucleon structure. Cross sections, electroweak (EW) parameters, coupling constants and scaling variables etc have also been measured by experimentalists through such processes. In the $\nu N$ DIS, the neutrino scatters off a quark in the nucleon via the exchange of a virtual W (CC) or Z (NC) boson, producing a lepton and hadronic system in the final state. Similarly, UHE neutrino cross sections have gained importance as many experiments worldwide are ongoing/planned to observe processes involving them. The natural sources of UHE could be - supernovae core collapse, cosmic rays, gamma ray burst, AGN etc and they serve as windows of understanding highest energy processes in the universe. Since attenuation of these neutrinos due to their travel is very low (as they are only weakly interacting) , they act as powerful tool to help us know about their sources. Various experiments measuring UHE neutrinos, ongoing and planned, worldwide are - Baikal \cite{4}, ANITA \cite{5}, RICE \cite{6}, AMANDA \cite{7}, HiRes \cite{8}, ANTARES \cite{9} , IceCube \cite{10}, GLUE \cite{10,11}, Pierre Auger Cosmic Ray Observatory \cite{12}, ARIANNA \cite{13}, JEM-EUSO \cite{14}. A number of studies on UHE neutrino cross sections (CC and NC) are available in literature. R. Gandhi, et al., \cite{3} (GQRS1998) reported results based on u,d,c,s quark PDFs (Parton Distribution Function) from 1998 CTEQ4 analysis of the early HERA-ZEUS small $\mathit{x}$ data. In the results presented by A. Connolly, et al., \cite {15} (CTW 2011) and A. Cooper-Sarkar, et al., \cite{16} (CSMS 2011) they included b-quark contribution to both CC and NC scattering and are based on updated PDFs obtained from newer data. Froissart bound inspired behaviour of $F_{2}^{ep}$ of DIS $(e-p)$ scattering was used by Martin M.Block, et al., \cite {17} (BDHM 2013) to evaluate UHE neutrino cross section off an isoscalar nucleon  $N = \frac{n+p}{2}$ , upto $E_\nu\sim10^{17}$ GeV. It may be noted that $E_\nu\sim10^{17}$ GeV is the highest reach of the experimental search for UHE cosmic neutrino \cite{10,11}. \par In this work, we calculate CC and NC neutrino-nucleon scattering cross section with $E_{\nu} \sim (10^{9} - 10^{12}$ GeV) using QCD inspired Double Asymptotic limit (DAL) of the parton structure function $F_{2}^{ep} (x, Q^{2})$.
The preliminary results of this analysis were presented in \cite{18}. In \cite{19} one of us found a form of dynamic pomeron type

\begin{equation}
F_{2}^{ep} \sim x^{-\lambda(Q^{2})},
\end{equation}
can be derived from DGLAP evolution equation, which was found to describe the available HERA H1 data for $F_{2}^{ep}$ in the range $ 1 \le x \le 10^{-4}$ and $5 \le Q^{2} \le 5000  GeV^{2}$ within 10 $\%$ error. It is worth mentioning that this behaviour Eq (1) of $F_{2}^{ep}$ can be viewed as of dynamic-pomeron type. In physics the pomeron  is a Regge trajectory, a family of particles with increasing spin, postulated to explain the slowly rising cross section of hadronic collisions at high energies. 
At high energies (and low $Q^{2}$) $\gamma^{\ast}p$  cross section is believed to have similarities to that of hadron-hadron interactions. Pomeron type behaviour of $F_{2}$ at small $\mathit{x}$ can explain the logarithmic rise of cross section with energy. In Figure 1 we present the result from the computation for CC, NC and total cross section for $10^{9} \le E_{\nu} \le 10^{12}$ GeV. 
We then compare our results (shown in Figures 2, 3) in the energy range $10^{9} \le E_{\nu} \le 10^{12}$ GeV with those already available in literature. While overall behaviour is found to be similar, the values of our cross sections are found to be lower than those of BDHM2013, CTW2011 and CSMS2011 for $E_{\nu} \ge (10^{9} - 10^{11})$ GeV for CC . On the other hand, our values are lower than those of GQRS1998, for energy of $E_{\nu} =10^{9}$ GeV. For NC, for $E_{\nu}\ge (10^{9} - 10^{12}$) GeV , our values are almost same as GQRS1998 whereas for $E_{\nu} \ge (10^{9} - 10^{10})$ GeV our values are slightly lower than BDHM2013, CTW2011 and CSMS2011. In our view, this could be attributed to the form of structure function (Eqn(1)), used to calculate $\nu N$ cross section. It may be noted that the rising behaviour of $F_{2}^{ep}$ can be controlled due to screening corrections and we intend to do it in our future work. Then we present analytical form of total cross section, fitted to a form, Eqns (25,26) both for CC and NC. The behaviour of Eqns (25,26) appears like a pomeron exchange, with an intercept $\alpha(0)\sim 0.48$. This looks like a hard pomeron, as explained in the text. We also fit our cross sections to the form Eqns (27,28) (as discussed in Eqn (30), obtained in \cite{27}), using the Reggeon diagram technique (RDT). \\It has been stated in \cite {1} that for a more accurate prediction of the $\nu N$ cross-section, a well formulated model of the nucleon structure function is needed and that this predictive power is specially important in the search of New Physics (NP). At such Ultra High Energies, the $\nu N$ cross section can depart substantially from the standard model predictions, if NP is at play. Study of such UHE neutrino interaction thus could be a possible probe of new physics. Determination/measurement of $\nu N$ cross section could also be useful to constrain the underlying QCD dynamics of the nucleon. Detection of UHE neutrino events may shed light on the observation of air shower events with energies $\geq 10^{10}$ GeV, as well. Moreover, the behaviour of UHE $\nu N$ cross section can also be used to discriminate among different models of gluon dynamics at play at very low $\mathit{x}$.
The energy dependence of total $\nu N$ cross section measurement may have important implications for hadronic interactions at such UHE, not accessible otherwise.  If cross section much outside the limits of ongoing/planned neutrino experiments are observed, then predictions presented in this work could be very important. This commands attention also, since many experiments worldwide are planned/ongoing in DIS/UHE regime.\\ We would like to emphasize here that, we have not used any software available in public domain, in our work we have done the computation of $\nu N$ cross section, using our own computer program and this is a novely in this work. Another novel feature is dynamic pomeron type behaviour of $F_{2}^{ep}$ used in our work, which also gives pomeron type behaviour for $\nu N$ cross section at UHE.  

The paper has been systematically organised as follows. In section II, we present a review on DAL behaviour of $F_{2}^{ep}$, following \cite {19} and its subsection II.A contains brief details about Regge theory. In section III, analytical formulae for $\sigma_{\nu N}^{CC}$ and $\sigma_{\nu N}^{NC}$, using above form of $F_{2}^{ep}$ are presented. Section IV contains numerical calculations, results and analysis. Lastly we summarize and draw conclusions in section V.
\section{A brief review of $F_{2}^{ep}(x,Q^{2})$ using DAL of QCD}
In this section, we describe briefly about electron-proton structure function $F_{2}^{ep}(x,Q^{2})$  utilising DAL of QCD, following \cite {19}, for the sake of completeness of this work. It is well known that in DIS $(e-p)$ scattering, the incoming electron scatters off the target proton, via the exchange of a virtual photon, producing a hadronic system in the final state. A typical $(e-p)$ DIS event can be described with the help of two independent variables, $\mathit{x}$ and $Q^{2}$, where $\mathit{x}$ is the Bjorken variable (fraction of proton’s momentum carried by its constituent partons, in Breit's frame) and $Q^{2}$ is the transverse momentum squared of the virtual exchanged photon. The scattering cross section can be described in terms of two structure functions, $F_{2}(x,Q^{2})$ and $F_{1}(x,Q^{2})$. Bjorken variable $x = \frac{Q^{2}}{2M\nu}$, here $\nu$ is the electron's energy loss and $Q^{2}$ depends on the scattering angle. The squared mass $W^{2}$ of the observed hadronic system is 
\begin{equation}
 W^{2} = (p+q)^{2} = M^{2}-Q^{2} + 2M\nu, 
\end{equation}
(in proton's rest frame)
where p and q are proton and electron's momentum respectively, M is proton's mass.  For elastic scattering, $W^{2} = M^{2}(x = 1)$. In parton model, at large $Q^{2}$, for spin $\frac{1}{2}$ partons,
 $F_{2}(x) = 2xF_{1}$ and $F_{L} = F_{2} - 2xF_{1} = 0$.

For point like parton, Bjorken scaling occurs, structure function do not depend on $Q^{2}$. But scaling violations are found to occur in $(e-p)$ DIS processes, as $\mathit{x}$ decreases, which means that structure function $F_{2}^{ep}$ depends on $Q^{2}$ also. Thus the proton no longer consists of point like partons only, but has a dynamic structure deep inside, which can be explained via QCD evolution equations in leading log $Q^{2}$ approximation ($LLQ^{2}$), known as DGLAP equations. 
In $(e-p)$ DIS, in the next to leading order, scaling violations occur through gluon bremsstrahlung from quarks and quark pair creation from gluons. At small $ x< 10^{-2}$, the latter process dominates the scaling violations. This property can be exploited to extract gluon density from the slope $\frac{dF_{2}}{dlnQ^{2}}$ of the proton structure function. The general equations \cite{14} describing the $Q^{2}$ evolution of the quark density and gluon density respectively are
\begin{equation}
\frac{dq^{i}(x,t)}{dt} = \frac{\alpha(t)}{2\pi}\int_{x}^{1}\frac{dy}{y}[\sum_{j=1}^{2f}q^{j}(y,t)P_{qq}(\frac{x}{y}) + G(y,t)P_{qG}(\frac{x}{y})],
\end{equation}
\begin{equation}
\frac{dG(x,t)}{dt} = \frac{\alpha(t)}{2\pi}\int_{x}^{1}\frac{dy}{y}[\sum_{j=1}^{2f}q^{j}(y,t)P_{Gq}(\frac{x}{y}) + G(y,t)P_{GG}(\frac{x}{y})],
\end{equation}
where $P_{qq}(\frac{x}{y}), P_{qG}(\frac{x}{y}), P_{Gq}(\frac{x}{y}), P_{GG}(\frac{x}{y})$ are the splitting functions and $t = ln\frac{Q^{2}}{Q_{0}^{2}}$. Assuming that the quark densities are negligible and the non-singlet contribution $F_{2}^{NS}$ can be ignored safely at small $\mathit{x}$ in DGLAP equation, for $F_{2}$, the equation becomes
\begin{equation}
\frac{dF_{2}(x, Q^{2})}{dlnQ^{2}} = \frac{10\alpha_{s}}{9\pi}\int_{x}^{1}dx^{'}P_{qG}(x^{'})\frac{x}{x^{'}}g(\frac{x}{x^{'}},Q^{2}).
\end{equation}
Here $xg(x,Q^{2}) = G(x, Q^{2})$ is the gluon momentum density and $g(x,Q^{2})$ is the gluon number density of the proton and $\frac{x}{x^{'}} = \frac{x}{y}$. Rearranging equation (5) we have 
\begin{equation}
\frac{dF_{2}(x, Q^{2})}{dlnQ^{2}} = \frac{5\alpha_{s}} {9\pi}\int_{x}^{1}dy\frac{x}{y}g(y, Q^{2})\frac{1}{y^{2}}[x^{2}+(y-x)^{2}].
\end{equation}
Substituting $y = \frac{x}{1-z}$ we can write RHS of equation (6) as 
\begin{equation}
\frac{5\alpha_{s}}{9\pi}\int_{0}^{1-x} dz G(\frac{x}{1-z}, Q^{2})[z^{2}+(1-z)^{2}].
\end{equation}
Expanding $G(\frac{x}{1-z},Q^{2})$ about $z = \frac{1-x}{2}$ and keeping terms upto the first derivative of G in the expansion we have
 
\begin{equation}
 G(\frac{x}{1-z},Q^{2})= G(\frac{2x}{1+x},Q^{2}) + (z-\frac{1-x}{2})\frac{4x}{(1+x)^{2}}\frac{dG(x^{''},Q^{2})}{dx^{''}} \pmb{\Bigg\vert}_{x^{''}=\frac{2x}{1+x}}.
\end{equation}
 
When this expansion is used in equation (6) we get 
\begin{equation}
\frac{dF_{2}(x,Q^{2})}{dlnQ^{2}} = \frac{5\alpha}{9\pi} \frac{(A+Ax+2B)^{2}}{(1+x)(A+Ax+4B)}G(y^{'},Q^{2}),
\end{equation}
where $y^{'} = [\frac{2x}{1+x}\frac{(A+Ax+4B)}{(A+Ax+2B)}]$, $A = [\frac{2(1-x)^{3}}{3}-(1-x)^{2}+(1-x)]$ and $B = [\frac{(1-x)^{4}-(1-x)^{3}}{6}]$ . In the limit $x\rightarrow 0$ equation (9) reduces to 
\begin{equation}
\frac{dF_{2}(x,Q^{2})}{dlnQ^{2}} = \frac{10\alpha_{s}}{9\pi} \frac{(1-x)^{2}}{(1-1.5x)} G (2x\frac{(1-1.5x)}{(1-x^{2})}, Q^{2}).
\end{equation}
Using the above, double asymptotic expression \cite{19} for $F_{2}$ in small $\mathit{x}$ and large $Q^{2}$ (DAL) limit, we can write
\begin{equation}
F_{2}^{p} \sim \frac{exp\sqrt{\frac{144}{33-2n_{f}}\xi ln(\frac{1}{x_{1}})}}{(\frac{144}{33-2n_{f}}\xi ln(\frac{1}{x_{1}}))^{\frac{1}{4}}},
\end{equation}
with $\xi = ln(\frac{ln\frac{Q^{2}}{\Lambda ^{2}}}{ln\frac{Q_{0}^{2}}{\Lambda ^{2}}})$, 
$x_{1} = \frac{2x-3x^{2}}{1-x^{2}}$ , $n_{f}$ is the number of flavors, $Q_{0}^{2}$ is the value at which the input parton parameterization is to be used and $\Lambda$ is the QCD mass scale. $F_{2}^{p}$ in equation (11) in DAL can be parametrized as 
\begin{equation}
F_{2}^{p} \sim x^{-\lambda(Q^{2})},
\end{equation}
which can be viewed as of dynamic pomeron type.
\subsection{Regge theory, Pomeron and Structure function $F_{2}^{ep}(x, Q^{2})$}
It was shown by Regge \cite{20} that non-relativistic potential scattering may be explained in a useful way, by considering angular momentum to be complex, and the simple pole type singularities in the scattering amplitude are called `Regge poles'. The eigenvalues of the corresponding Schrodinger's equation lie in a trajectory, called Regge trajectory
\begin{equation}
l = \alpha (t),
\end{equation}
where $\mathit{t}$ is the square of centre of mass energy in $\mathit{t}$- channel. The basic idea was that sequences of composite particles of mass $m_{i}$ and spin $\sigma_{i}$ would lie on a given Regge trajectory such that 
\begin{equation}
\alpha (m_{i}^{2}) = \sigma_{i}.
\end{equation}
In Regge theory, the high energy behaviour of scattering amplitude A(s,t) can be explained by exchange of trajectory of particles
\begin{equation}
A(s,t) \sim s^{\alpha(t)},
\end{equation}
and the total cross section behaves as 
\begin{equation}
\sigma_{tot} (s) \sim s^{\alpha (0) - 1},
\end{equation}
where $\alpha (t) = \alpha (0) + \alpha^{'} t$.
Equation (16) is true upto leading order in $\mathit{s}$, and corrections of order $s^{\alpha (t) - 1}$ may be anticipated due to daughter trajectories, threshold corrections etc. 
A trajectory called Pomeron (or Pomerenchuk) with intercept 
\begin {equation}
\alpha_{p}(0) \approx 1,
\end{equation}
was invented \cite{21,22} to account for the asymptotic behaviour of total cross section for hadron-hadron scattering at large energies and $\alpha_{p}(0) \approx 1$ was conjectured to be the maximum value permitted by the Froissart bound of cross section. But rising cross section with energy and complications of pomeron cuts make one wonder if the pomeron may be a more complicated singularity than a pole \cite{20}. When applied to $(e-p)$ scattering and behaviour of structure function $F_{2}^{ep}$, Regge theory is expected to be applicable at high energies and small $\mathit{x}$, even for large values of $Q^{2}$. Until HERA measurements of $F_{2}^{ep}$, generally a soft pomeron (applicable at low $Q^{2}$) with intercept close to 1.08 \cite{23,24,25} seemed to describe the small $\mathit{x}$ structure function data. As discussed above, Regge theory relates the W-dependence of the process to the position of singularities in complex angular momentum plane and the relative strenghts of the contributions from these singularities may vary with $Q^{2}$. The HERA data seemed to exhibit a pomeron-type behaviour with higher intercept \cite{23} and idea of two pomerons - one soft and another hard (for large $Q^{2}$) has been explored \cite{23}.
\begin{equation}
F_{2}^{ep} (x, Q^{2}) = \Sigma f_{i}(Q^{2})x^{-\epsilon_{i}},
\end{equation}
where $\epsilon_{1} = 0.0808$ (soft pomeron) and $\epsilon_{2} = 0.418$ (hard pomeron).
It has been argued that the pomeron may be just glueball trajectories -the soft pomeron was conjectured to lie on a $2^{++}$ glueball trajectory \cite{25,26} and that there may be glueballs at different masses. We would like to mention here that glueballs are hypothetical, composite particles that consist only of gluons. They are extremely difficult to identify in particle accelerators as they can mix with meson states. So far no signatures of glueballs in experiments have been found. Many new experiments worldwide, like FAIR \cite{28} at Germany, the GlueX experiment at Jefferson Laboratory in USA \cite{29} and ALICE at CERN \cite{30} may provide us some signature in future on glueballs. 

\section{Total charged and neutral current neutrino nucleon cross section at UHE}

The total charged and neutral current (CC and NC) for neutrino nucleon scattering \cite{17}, for an isoscalar nucleon N = $\frac{n+p}{2}$, can be written as 
\begin{equation*}
\sigma^{\nu N}_{CC}(E_{\nu}) = \int_{Q_{min}^{2}}^{s} dQ^{2 }\int_{\frac{Q^2}{s}}^{1}dx \frac{d^{2}\sigma_{CC}}{dx dQ^{2}}(E_{\nu}, Q^{2}, x)
\end{equation*}
\begin{equation}
= \frac{G_{F}^{2}}{4\pi}\int_{Q_{min}^{2}=1}^{2mE_{\nu}} dQ^{2}(\frac{M_{W}^{2}}{Q^{2}+M_{W}^{2}})^{2}\int_{\frac{Q^{2}}{2mE_{\nu}}}^{1}\frac{dx}{x}\bigg\lbrack F_{2}^{\nu} + x F_{3}^{\nu} + (F_{2}^{\nu}-x F_{3}^{\nu})\Big(1-\frac{Q^{2}}{xs}\Big)^{2}-\Big(\frac{Q^{2}}{xs}\Big)^{2} F_{L}^{\nu}\bigg],  
\end{equation}
where $F_{2}^{\nu}$ is the neutrino-nucleon structure function, $\mathit{s} = 2mE_{\nu}$, where $\mathit{s}$ is the Mandelstam variable which is the total energy in the centre of mass frame, $\mathit{m}$ is the nucleon mass, $G_{F}$ is the Fermi constant and $M_{W}^{2}$ is the squared mass of intermediate W-boson, $Q^{2}$ is the four momentum square of virtual photon. Here $\mathit{xF_{3}}$ is a measure of difference of quarks and antiquarks PDFs, and so is sensitive to the valence quark distribution function. We neglect valence quark contribution in our analysis, as at small $\mathit{x}$, structure of proton is dominated by gluons only \cite{19}. Therefore, contributions of $F_{3}^{\nu}$ to $\nu N$ scattering is sub dominant only and hence neglected in our analysis. Similar expression can be obtained for neutral current cross section by replacing $M_{W}$ by $M_{Z}$ in Eqn (19). For the flavor-symmetric ($q\bar{q}$)N interaction at small $\mathit{x} < 0.1$, the neutrino-nucleon structure function, $F_{2}^{\nu}(x, Q^{2})$ can be related to electromagnetic structure function, $F_{2}^{ep}(x, Q^{2})$ \cite {31} as 
\begin{equation}
F_{2}^{\nu}(x, Q^{2}) = \frac{n_{f}}{\sum_{q}^{n_{f}}Q_{q}^{2}} F_{2}^{p}(x, Q^{2}),
\end{equation}
where $n_{f}$ is the number of flavors and $Q_{q}$ is the quark charge.
Thus for $10^{9}<E_{\nu}<10^{12}$, $\mathit{x}$ lies in the range $10^{-5}<x<10^{-8}$. Here
$\sigma_{CC,NC}^{\nu N}$, is the neutrino nucleon cross section - to leading order in weak coupling $G_{F}$ and all orders in strong hadronic interaction.\\
Minimum value of $Q^{2}$ is consistant with application of pQCD, we have used $Q^{2}_{min}$ = 1 $GeV^{2}$ in our computation. Now using DAL value of $F_{2}^{ep}$ from Eqn (12) in Eqn (19), we obtain the expression for total neutrino-nucleon cross sections as  
\begin{equation}
\sigma^{\nu N}_{CC}(E_{\nu})\approx \frac{G_{F}^{2}}{4\pi}\int_{Q_{min}^{2}=1}^{{2mE_{\nu}}\times10^{-2}} dQ^{2}(\frac{M_{W}^{2}}{Q^{2}+M_{W}^{2}})^{2}\int_{\frac{Q^{2}}{2mE_{\nu}}}^{1}\frac{dx}{x}(x^{-\lambda(Q^{2})}),
\end{equation}
where $\lambda(Q^{2}) = a - b.e^{-cQ^{2}}$ and the values of constants are found to be as $a = 0.486 , b = 0.272$ and $c = 0.002$. Solving Eqn (21) we get  
$$\sigma_{CC}^{\nu N}(E_{\nu}) = A \frac{G_{F}^{2}}{4\pi}\int_{Q_{min}^{2}=1}^{{2mE_{\nu}}\times10^{-2}} \frac{dQ^{2}}{-\lambda(Q^{2})}(\frac{M_{W}^{2}}{Q^{2}+ M_{W}^{2}})^{2} {x^{-\lambda(Q^{2})}}\pmb{\Bigg\vert}_{\frac{Q^{2}}{2mE_{\nu}}}^{1}$$ 
\begin{equation}
= - A \frac{G_{F}^{2}}{4\pi}\int_{Q_{min}^{2}=1}^{{2mE_{\nu}}\times10^{-2}} \frac{dQ^{2}}{\lambda(Q^{2})}(\frac{M_{W}^{2}}{Q^{2}+ M_{W}^{2}})^{2} \{1-(\frac{Q^{2}}{2mE_{\nu}})^{-\lambda(Q^{2})}\} 
\end{equation}
\begin{equation}
=  A \frac{{G_{F}^{2}}M_{W}^{4}}{4\pi}\int_{Q_{min}^{2}=1}^{{2mE_{\nu}}\times10^{-2}} \frac{dQ^{2}}{\lambda(Q^{2})}(\frac{1}{Q^{2}+ M_{W}^{2}})^{2} \{{(\frac{Q^{2}}{2mE_{\nu}})^{-\lambda(Q^{2})}-1}\},
\end{equation}
in low $\mathit{x}$ and high $Q^{2}$ regime. Here A is normalisation constant.\\

The corresponding total neutral current cross section $\sigma_{NC}^{\nu N}(E_{\nu})$ is obtained by replacing $M_{W}$ by squared mass of intermediate Z boson $M_{Z}$ that is 
\begin{equation}
 \sigma_{NC}^{\nu N}(E_{\nu}) =  A \frac{{G_{F}^{2}}M_{Z}^{4}}{4\pi}\int_{Q_{min}^{2}=1}^{{2mE_{\nu}}\times10^{-2}} \frac{dQ^{2}}{\lambda(Q^{2})}(\frac{1}{Q^{2}+ M_{Z}^{2}})^{2} \{{(\frac{Q^{2}}{2mE_{\nu}})^{-\lambda(Q^{2})}-1}\}.
\end{equation}

\section{Results and Discussion}

We have computed $\sigma_{CC}^{\nu N}$ and $\sigma_{NC}^{\nu N}$ by carrying out the integration in Eqns (23, 24) using our own computation (Monte Carlo integration technique) and have presented the results in Figure 1. We find that the behaviour of $\sigma_{CC}^{\nu N}$ and $\sigma_{NC}^{\nu N}$ is similar to that available in literature. The values of our total cross section both for charged and neutral current are organised in tabular form (Tables 1 and 2) along with other cross section values that are available in literature. Also values of tables 2 and 3 are presented in Figures 2 and 3 respectively. We then make a fit to the CC and NC $\nu N$ cross sections to obtain the analytic forms of the following types in the energy range $10^{9} GeV \le E_{\nu} \le 10^{12} GeV $ : 
\begin{equation}
 \sigma_{CC}^{\nu N} = (-31.3351 \pm 1.11623) \times E_{\nu}^{(-0.423249 \pm 0.0176539)} + (-20.054 \pm 0.481768), 
 \end{equation}
 \begin{equation}
 \sigma_{NC}^{\nu N} = (-31.4594 \pm 1.10901) \times E_{\nu}^{(-0.424348 \pm 0.0174823)} + (-20.1152 \pm 0.477422).
\end{equation}

 This can be viewed as a hard-pomeron (with intercept $\alpha(0)\sim 0.48$) type behaviour of the cross section at UHE. Here we would like to emphasize that, a dynamic pomeron type form of $F_{2}^{ep}$ (Eqn (1)), of the strong interactions, gives a hard-pomeron type behaviour of (Eqns 25, 26) total cross section of weak interactions. This could point out to some interplay between strong and weak interactions. It is well known that due to higher order corrections, coupling constants of all the gauge interactions change with energy scale, and are believed to unify at higher scale $M_{GUT} \sim 2 \times 10^{16}$ GeV in Grand Unified theories (GUTs). So, the electroweak coupling constant increases, while the coupling constant of strong interaction decreases as energy increases. Therefore, a neutrino may behave as a strong particle, as far as its coupling constant is concerned, at higher enough scales $\geq 10^{12}$ GeV, and hence the ($\nu-p$) scattering cross section may have a similiar behaviour as hadron-hadron scattering cross section. This justifies the pomeron type behaviour of ($\nu-p$) scattering cross section at high energies. We also fit our cross section to the form 
\begin{equation}
\sigma_{CC}^{\nu N} = (- 32.2999 \pm  0.062563) + (-4.45714 \pm 0.035448) \times ln E_\nu + (2.00921 \pm 0.0113361) \times (ln E_{\nu})^{2},
\end{equation}
\begin{equation}
\sigma_{NC}^{\nu N} = (- 32.3876 \pm  0.057913) + (-4.46797 \pm 0.032813) \times ln E_\nu + (2.0157 \pm 0.010493) \times (ln E_{\nu})^{2}.
\end{equation} 
 In perturbative QCD, the ladder of gluons may be thought to behave as a pomeron (reggeon)  if one uses reggeon diagram technique \cite{27}. Here, the hadron-hadron scattering is explained in terms of leading log diagrams (as discussed earlier), and a connection can be established in asymptotic behaviour and $\mathit{t}$-channel singularity. 
The asymptotic behaviour of $\sigma_{t}$ can be described as \cite{27}
$$\sigma_{t} \sim \frac{1}{\surd{lns}} e^{(2ln s.ln 2)}$$
\begin{equation}
\sim \frac{s^{2ln2}}{\surd{ln s}},
\end{equation}
which also leads to the well known Froissart bound on cross section
\begin{equation}
\sigma_{t} \sim (ln s)^{\eta}, \qquad  \eta \leq 2.
\end{equation}
We would like to mention here that the rise in $\sigma_{t}$ as $\mathit{s\rightarrow} \infty$, (as observed in our results) can be controlled with the help of screening corrections and this study will be done in our future work \cite{32}.
\section{Summary}
To summarize, in this work, we have calculated the pomeron inspired behaviour of $\nu N$ cross section in Ultra High Energy limit. We presented a brief review of $F_{2}^{ep}$ using DAL of QCD. Then we have calculated total neutrino-nucleon cross section $\sigma_{\nu N}^{CC}$ for CC and $\sigma_{\nu N}^{NC}$ for NC interactions using the Double Asymptotic Limit of $F_{2}^{ep}$ (which resembles dynamic pomeron type behaviour) of DIS ($\textit{e-p}$) scattering, found earlier by one of us \cite {19}. In \cite {3}, \cite {15,16,17}, they used standard sets of parton distribution functions available in literature at that times, to obtain total cross sections at UHE, but we have used our own parameterization for $F_{2}^{ep}$ (within 10 $\%$ error) in DAL, using input PDFs at $Q_{min}^{2}$. We used Monte Carlo integration technique in our computation to obtain these cross sections in the energy range  $10^{9} GeV \le E_{\nu} \le 10^{12} GeV $. Then we did a parameter fitting of these cross sections, to obtain their analytical form (Eqns 25 - 28). We found that though the overall behaviour of our calculated $\nu N$ cross sections is similar to the above mentioned works, our values are slightly smaller, in the low energy range, while larger in the high energy range. This difference could be attributed to different assumptions in input parameterization of PDFs used in $F_{2}^{ep}$, and due to the fact that we have used our own analytic form of $F_{2}^{ep}$ in low $\mathit{x}$ and large $Q^{2}$ regime obtained from DGLAP equation. 
So, the dynamic pomeron type form of nucleon structure function $F_{2}^{ep}$ gives lower charged current $\nu N$ cross section in lower energy ($E_{\nu} \leq 4 \times 10^{10}$ GeV) and also slightly higher $\nu N$ cross section in higher energy ($E_{\nu} \geq 2 \times 10^{10}$ GeV). This could be attributed to the pomeron type behaviour of $F_{2}^{ep}$ which gives a higher slope of $\nu N$ cross section. The dynamical pomeron-type behaviour of $F_{2}^{ep}$ give rise to a hard-pomeron (with intercept $\alpha(0) \sim 0.48$) type behaviour of total neutrino cross section in UHE regime. This could hint to some interplay between strong ($F_{2}^{ep}$) and weak ($\sigma^{\nu N}$) dynamics. The future measurements of $\sigma^{\nu N}$ in this regime would provide a test to the ideas presented in the work. 

\section{Acknowledgements}
Bora.K and Sarma.N would like to thank DST-SERB, Govt of India, for a project; Grant No.DST-SERB/EMR/2014/000296 under which this work is done. N.Sarma also thanks Prof. Raj Gandhi, for financial support and useful discussions at Harish Chandra Research Institute (HRI), Allahabad, India, where a part of this work has been done. Singh.J is grateful to HRI, Allahabad.

\begin{figure}[htp]
\includegraphics[scale= .45]{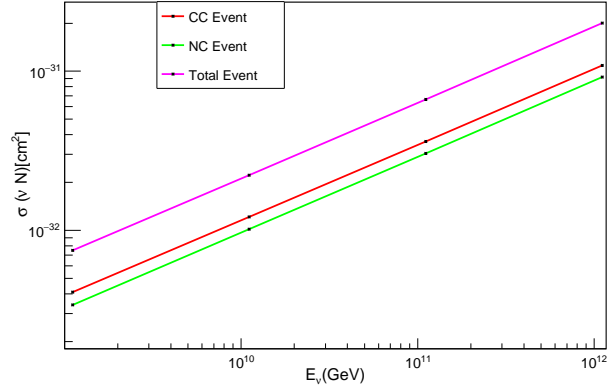}
\caption{Variation of neutrino-nucleon charged current, neutral current and total current cross sections with neutrino energy (from our calculation).}
\end{figure}

\begin{figure}[htp]
\includegraphics[scale= .45]{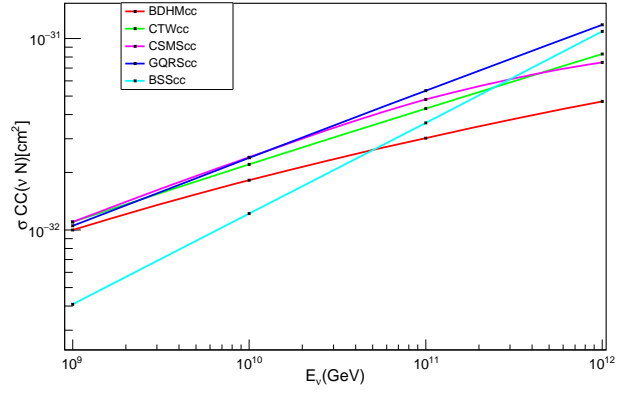}
\caption{Comparison of charged current $\nu N$ cross sections, in $cm^{2}$ as a function of $E_{\nu}$.}
\end{figure}

\begin{figure}[htp]
\includegraphics[scale= .45]{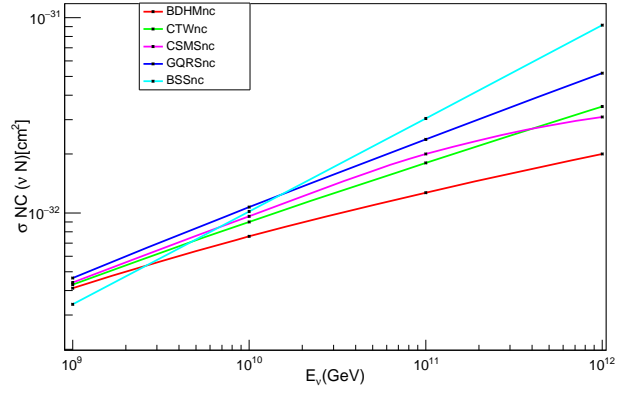}
\caption{Comparison of neutral current $\nu N$ cross sections, in $cm^{2}$ as a function of $E_{\nu}$.}
\end{figure}

\begin{table}[htp]
\caption{Charged current $\nu N$ cross sections, in $cm^{2}$ as a function of $E_{\nu}$ are listed. Here BDHM refers to the work done by Martin M.Block, et al., \cite{17}, CTW refers to A. Connolly,  et al., \cite{15}, CSMS refers to A. Cooper-Sarkar, et al., \cite{16}, GQRS refers to R. Gandhi, et al., \cite{3} and BSS refers our work in this paper.}
\renewcommand\thetable{\Roman{table}}
\centering
\setlength{\tabcolsep}{2pt}
\begin{tabular}{|c | c | c | c | c | c |}
\hline
$E_{\nu}$ (GeV) & $\sigma_{BDHM}$($cm^{2}$) & $\sigma_{CTW}$($cm^{2}$) & $\sigma_{CSMS}$($cm^{2}$) & $\sigma_{GQRS}$($cm^{2}$) & $\sigma_{BSS}$($cm^{2}$)\\
\hline\hline

$10^9$  &    $1.00\times10^{-32}$  &  $1.1\times10^{-32}$  &  $1.1\times10^{-32}$  &   $1.05\times10^{-32}$    &  $4.09\times10^{-33}$ \\
$10^{10}$  & $1.82\times10^{-32}$  &  $2.2\times10^{-32}$  &  $2.4\times10^{-32}$  &   $2.38\times10^{-32}$    &  $1.21\times10^{-32}$ \\
$10^{11}$  & $3.02\times10^{-32}$  &  $4.3\times10^{-32}$  &  $4.8\times10^{-32}$  &   $5.34\times10^{-32}$    &  $3.62\times10^{-32}$\\
$10^{12}$  & $4.69\times10^{-32}$  &  $8.3\times10^{-32}$  &  $7.5\times10^{-32}$  &   $1.18\times10^{-31}$    &  $1.08\times10^{-31}$ \\ [1ex]
\hline
\end{tabular}
\end{table}

\begin{table}[htp]
\setlength{\tabcolsep}{2pt}
\caption{Neutral current $\nu N$ cross sections, in $cm^{2}$ as a function of $E_{\nu}$ are listed. Here BDHM refers to the work done by Martin M.Block, et al., \cite{17}, CTW refers to A. Connolly,  et al., \cite{15}, CSMS refers to A. Cooper-Sarkar, et al., \cite{16}, GQRS refers to R. Gandhi, et al., \cite{3} and BSS refers our work in this paper.}
\label{table:second}
\centering

\begin{tabular}{|c | c | c | c | c | c|}
\hline
$E_{\nu}$ (GeV) & $\sigma_{BDHM}$($cm^{2}$) & $\sigma_{CTW}$($cm^{2}$) & $\sigma_{CSMS}$($cm^{2}$) & $\sigma_{GQRS}$($cm^{2}$) & $\sigma_{BSS}$($cm^{2}$)\\
\hline\hline
$10^9$  &   $4.12\times10^{-33}$  &  $4.3\times10^{-33}$  &  $4.4\times10^{-33}$  &   $4.64\times10^{-33}$    &  $3.40\times10^{-33}$ \\
$10^{10}$  & $7.58\times10^{-33}$  &  $9.0\times10^{-33}$  &  $9.6\times10^{-33}$  &   $1.07\times10^{-32}$    &  $1.016\times10^{-32}$ \\
$10^{11}$  & $1.27\times10^{-32}$  &  $1.8\times10^{-32}$  &  $2.0\times10^{-32}$  &   $2.38\times10^{-32}$    &  $3.042\times10^{-32}$\\
$10^{12}$  & $2.00\times10^{-32}$  &  $3.5\times10^{-32}$  &  $3.1\times10^{-32}$  &   $5.20\times10^{-32}$    &  $9.16\times10^{-32}$ \\ [1ex]
\hline
\end{tabular}
\end{table}

\end{document}